\documentclass[pss]{wiley2sp} 
\usepackage{amsmath}

\begin{document}
\hyphenation{con-duc-tion-band-struc-tur-es}

\title{InP-quantum dots in Al$_{0.20}$Ga$_{0.80}$InP 
with different barrier configurations}

\titlerunning{InP-quantum dots in AlGaInP }

\author{%
  Wolfgang-Michael Schulz, 
  Robert Ro\bf{$\ss$}bach, 
  Matthias Reischle,
  Gareth~J.~Beirne,
  Michael Jetter, 
  and Peter Michler
  }

\authorrunning{W.-M. Schulz et al.}

\mail{e-mail:
  \textsf{m.schulz@ihfg.uni-stuttgart.de}, Phone:
  +49-711-685-69852, Fax +49-711-685-63866}

\institute{%
  Institut f\"ur Halbleiteroptik und Funktionelle Grenzfl\"achen, Allmandring 3, 70569 Stuttgart, Germany
  }

\received{XXXX, revised XXXX, accepted XXXX} 
\published{XXXX} 

\pacs{ } 

\abstract{%
%
%
%
\abstcol{%
  Systematic ensemble photoluminescence studies have been performed on type-I InP-quantum dots in Al$_{0.20}$Ga$_{0.80}$InP barriers, emitting at approximately 1.85~eV at 5~K. 
The influence of different barrier configurations as well as the incorporation of additional tunnel barriers on the optical properties has been investigated.
The confinement energy 
between the dot barrier and the surrounding barrier layers, which is the sum of the band discontinuities for the valence and the conduction bands, was chosen to be approximately 190~meV by using Al$_{0.50}$Ga$_{0.50}$InP. In combination with 2~nm thick AlInP tunnel barriers, the internal quantum efficiency of these barrier configurations can be increased by up to a factor of 20 at elevated temperatures with respect to quantum dots without such layers.}}
%
%
\titlefigure[height=5.1cm]{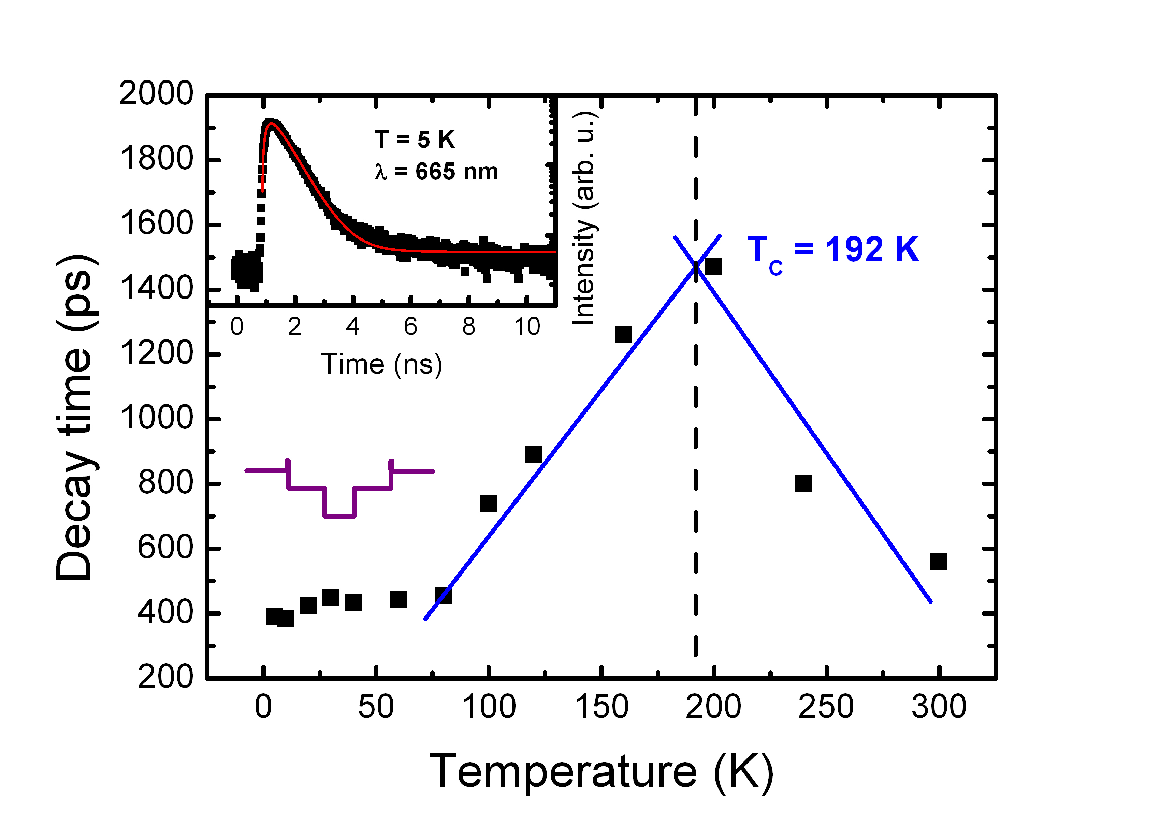}
\titlefigurecaption{%
Ensemble-transient (inset) and temperature dependent decay times of InP-QDs in Al$_{0.20}$GaInP-barriers with surrounding Al$_{0.50}$GaInP layers and AlInP tunnel barriers (sketch).}

\maketitle   
\section{Introduction}
Future quantum information applications with photonic devices requires high single-photon repetition rate emitters and detectors with high detection efficiencies. 
The highest sensitivity of a state-of-the-art ava--lanche-photodiode (APD) is nowadays in the red spectral range. These wavelengths are easily accessible by InP quantum dots (QDs) in (Al)GaInP-barriers. To make these emitters valuable for commercial applications, they should work at elevated temperatures. One possible solution is a deeper confinement potential for the QD to inhibit the thermal escape of the carriers.
In our previous work we have shown that the maximum luminescence intensity can be reached for InP-QDs in Al$_{0.20}$Ga$_{0.80}$InP-barriers especially at elevated temperatures \cite{rossbach}. The QD formation processes depend on the Al content of the barrier and thereby also strong--ly influence the optical properties of the QDs. In addition, the luminescence intensity temperature stability is strongly enhanced by the increased confinement energy, which can be normally attributed to the energy difference between the QD emission energy and the barrier emission energy. To further increase the carrier confinement and thus the luminescence intensity temperature stability, the QDs were also embedded in different barrier layers (claddings). 
The investigation of different barrier configurations also provides a deeper insight into the electrical behavior of such structures when embedded in p-i-n-diode architectures for future light and laser applications, as the carrier injection (leakage current, threshold current,..) will depend on these layers.
\section{Experimental procedure}
The self-assembled InP-QDs were grown by metal-organic vapor-phase epitaxy (MO\-VPE) with standard sources (trimethylgallium, tri\-me\-thyl\-indium, trimethylaluminum, arsine and phosphine) at low pressure (100 mbar) on (001) GaAs substrates oriented by 6$^\circ$ towards (111)A. The reference structure consists of a 100~nm GaAs buffer layer grown at 750$^\circ$C and a 100~nm thick lattice matched (Al$_{0.20}$Ga$_{0.80}$)$_{0.51}$In$_{0.49}$P layer grown at 720$^\circ$C. 
For the QD growth we deposited 2.1 monolayers (ML) of InP at a growth rate of 1.05~ML/s and a growth temperature of 650$^\circ$C. The relatively low growth temperature was chosen in order to inhibit Al incorporation from the barrier into the QDs during their self-assembly \cite{rossbach}. Nevertheless, the absolute aluminum content in the dots is unknown.
After a growth interruption of 20~s to ripen the QDs, a 30~nm cap-layer of Al$_{0.20}$Ga$_{0.80}$InP followed in the case of samples for photoluminescence (PL) measurements.  
For the barrier configuration series the total sample-thickness (cap and barrier thickness) was kept constant. The QD active region consists of symmetric, 10~nm thick Al$_{0.20}$\-Ga$_{0.80}$InP barriers. We kept the growth conditions of the active regions constant during the cladding series. Al$_{0.50}$\-Ga$_{0.50}$InP cladding was used to fabricate both the abrupt and gradual interfaces. 
This gives an additional total band offset of E$_{Off}$= 190~meV with respect to the barrier \cite{bornstein}. In addition to the integration of 2~nm thick tunnel barriers, the combination of an abrupt Al$_{0.50}$Ga$_{0.50}$InP cladding with 2~nm thick AlInP tunnel barriers, providing an additional total band offset of 60~meV, was examined.\\
The samples were placed in a cold finger liquid He-flow cryostat to vary the ensemble PL measurement temperature. The samples were excited using frequency doubled Ti:sapphire-laser emission for both the time-resolved and time integrated PL. The luminescence was then dispersed using a 0.32~m Jobin-Yvon monochromator and detected by a multi-channel plate photomultiplier tube.\\
In order to perform micro-PL ($\mu$-PL) experiments at 5 K, the samples were mounted in a liquid He-flow cryostat that can be scan\-ned both horizontally and vertically using two stepper motors with an effective spatial resolution of 50~nm. The light to and from the QDs was transmitted through a 50$\times$ microscope objective resulting in a laser spot diameter of approximately 1~$\mu$m when focused accurately using a piezo-based actuator. The PL was dispersed
using a 0.75~m spectrometer and detected by a liquid-nitro\-gen cooled charge-coupled device camera when taking spectra. For the photon statistics measurements, two single-photon counting avalanche photodiodes (APDs) were used, one in each path of a Hanbury–Brown and Twiss type
setup realised using a 50/50 beamsplitter after the output of the monochromator \cite{beirne2007,hbt56}.
\section{Results}
\subsection{Ensemble measurements}
\begin{figure}[Htbp]
\includegraphics[width=\linewidth]{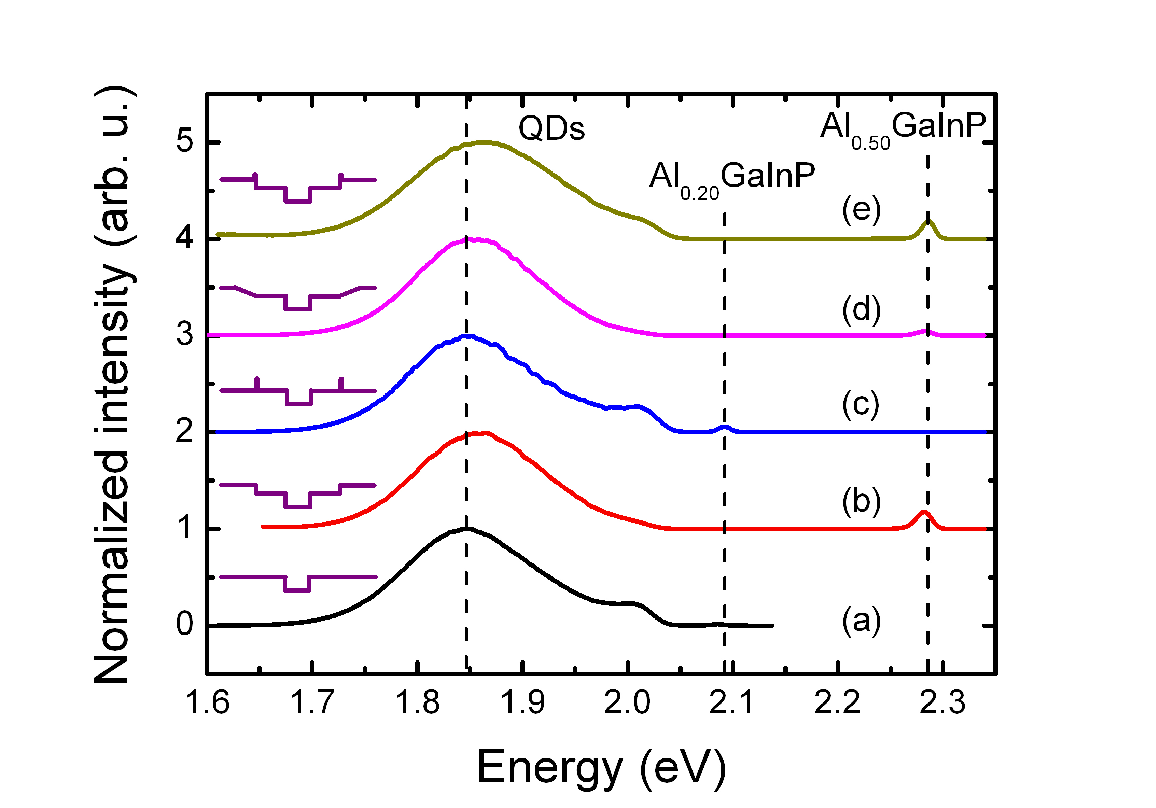}
\caption{PL spectra of InP-QDs in Al$_{0.20}$GaInP barriers with different barrier configurations, excited at 400~nm, 700~Wcm$^{-2}$ at 5~K. The growth and therefore the emission energy of the QDs, which is marked by the dashed vertical line at 1.847~eV, is not significantly influenced by the claddings. The other two vertical lines mark the emission of the Al$_{0.20}$GaInP barrier at 2.09~eV and the Al$_{0.50}$GaInP material at 2.28~eV. Emission from the AlInP tunnel barrier, which is expected at 2.33~eV, is not observed.} \label{Barrierenserie-vgl}
\end{figure}

\begin{figure}[Hbp]
\begin{centering}
\includegraphics[width=0.9\linewidth]{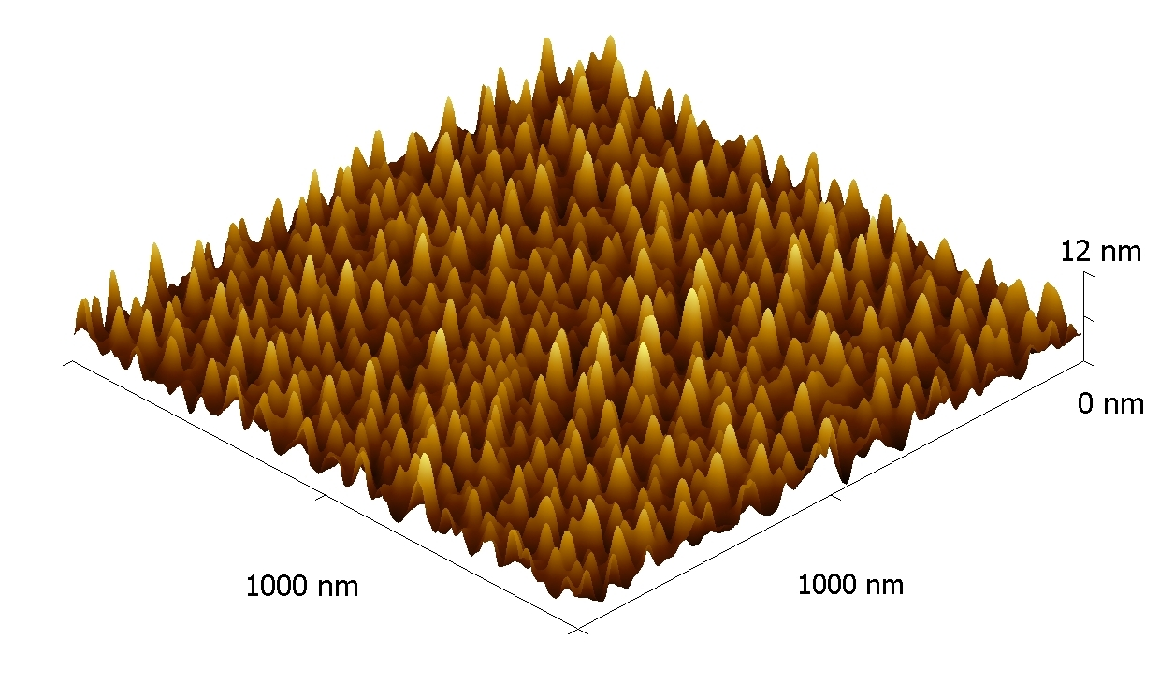}
\caption{AFM height image of open (uncapped) InP-quantum dots on an Al$_{0.20}$GaInP barrier (image size 1$\times$1~$\mu$m$^2$). The average height is approximately 4.3~nm (zero error subtracted) and the diameter is approximately 30~nm.
} \label{AFM}
\end{centering}
\end{figure}

\begin{figure}[h]
\includegraphics[width=\linewidth]{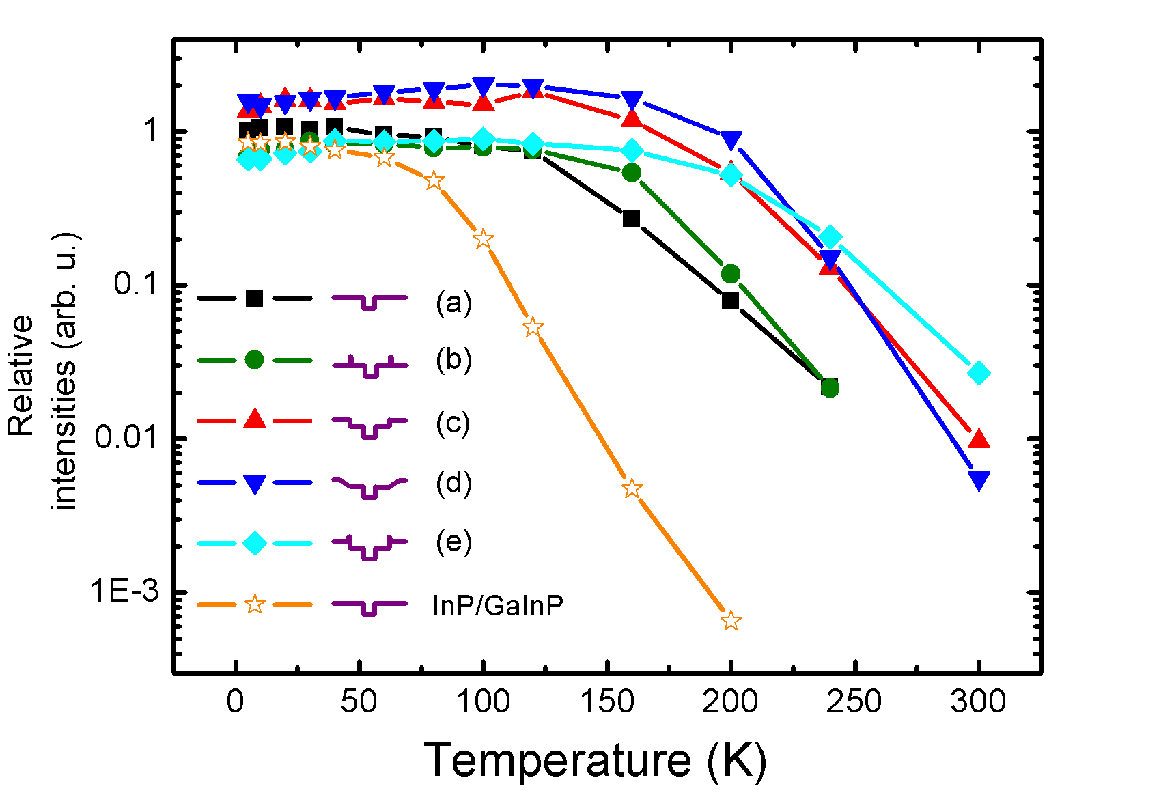}
\caption{Temperature dependence of the relative intensities of InP-QDs in Al$_{0.20}$GaInP barriers with different barrier configurations. For comparison the temperature dependence of InP-QDs in GaInP without cladding layers is also presented.} \label{IQEundRelintens}
\end{figure}

\begin{figure}[htbp]
\includegraphics[width=\linewidth]{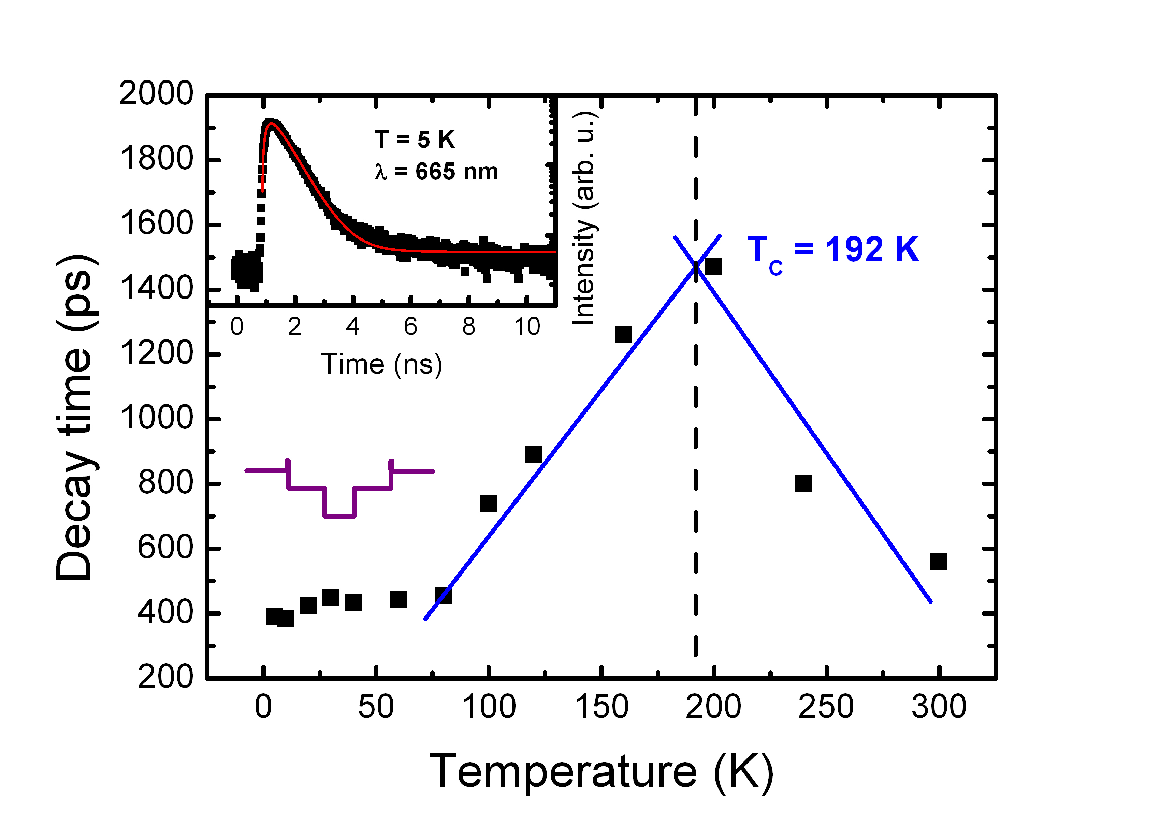}
\caption{Experimental decay time and transient (inset) of InP-QDs in the abrupt cladding structure with tunnel barriers at the dot center wavelength ($\lambda$= 665~nm) at 5~K. The inset shows the monoexponential decay, while the red curve is the fit derived from the model \cite{mukai}.} \label{M4220time}
\end{figure}

The normalized spectra and sketches of their conduction-band\-struc\-tures for the samples with different barrier configurations are shown in Fig. \ref{Barrierenserie-vgl}. 
The lowest spectrum (a) in the graph corresponds to the reference. One can see, that the different configurations have almost no influence on the spectral position and the full width half maximum of the QD emission.
The only change in the spectra (b), (d), (e) is the weak emission from the Al$_{0.50}$GaInP observed at around 2.28~eV. Emission from the AlInP tunnel barrier is not observed. As the emission characteristics of QDs are strongly influenced by their size, we conclude that there is also no significant change in dot height and its distribution. 
Therefore the confinement energy E$_C$, which is the energetic difference between QD and barrier emission, remains almost constant at a value of approximately E$_C$= 240 meV. 
Fig. \ref{AFM} shows an atomic force microscopy (AFM) height image of InP-QDs grown on an Al$_{0.20}$Ga\-InP barrier.
The QDs exhibit an average height of around 4.3~nm, with a sample surface density of approximately 5~$\times$10$^{10}$cm$^{-2}$.
The major difference between the structures (insets) and the influence of the cladding on the emission characteristics can be seen in the temperature behavior of their relative intensities as depicted in Fig. \ref{IQEundRelintens}. 
For comparison the temperature behavior for InP-QDs in GaInP without cladding (open stars) is also shown in this graph. At 200~K the InP-QDs in Al contaning barriers show a PL intensity enhancement by more than a factor 100 with respect to QDs, emitting at the same energy, in GaInP. According to Fig. \ref{IQEundRelintens}, the tunnel barriers (b) slightly reduce the intensity with respect to the reference structure (a). This is expected as they should reduce the number of excited carriers compared to the other cladding samples at comparable excitation powers. The highest intensity at low temperatures can be achieved using the graded barrier configuration (d), and is probably due to the associated gradient force that acts on the carriers. From a very basic point of view, this force $F$ can be written as $\vec{F}=-q\cdot \vec{\nabla} \Phi$, the gradient of the potential $\Phi$ (band offset between the barrier and the cladding) multiplied by the carrier charge $q$ \cite{nolting}. The carriers were pushed towards the QDs as this gradient force points in the direction of the local potential minimum. For this cladding configuration, the aluminum content of the barrier was gradually increased from 20$\%$ to 50$\%$ within a thickness of 10~nm. But as the grading forms a smooth transition, the carriers can be lost more easily at elevated temperatures because the phonon-carrier scattering and therefore the escape of carriers is not restricted to distinct phonon energies. Therefore the use of abrupt Al$_{0.50}$GaInP claddings in combination with thin AlInP tunnel barriers (e) is recommended for high luminescence intensities at elevated temperatures. 
The temperature dependent PL intensity was fitted using a simple thermal activation model \cite{sugisaki2002} to estimate the activation energy E$_A$.
In addition, time-resolved PL measurements were performed on the structures. As an example in the inset of Fig. \ref{M4220time} the transient of the PL maximum at 5~K can be seen for the structure with the abrupt cladding and AlInP tunnel barriers. The monoexponential behavior can be described using a model of the random initial occupation \cite{mukai} to determine the decay times.  
The QDs display an approximately constant decay time (Fig. \ref{M4220time}) up to 80~K with a linear increase at higher temperatures \cite{citrin}. 
For elevated temperatures above 200~K the decay time decreases due to excitonic dissociation to the barrier. We marked the intersection between the two fitted lines where the escape of the carriers into the barrier begins to dominate. This characteristic temperature T$_C$ was estimated for all structures. 

Table \ref{comp} summarizes the activation energies and the characteristic temperatures for the different structures under investigation. 
\begin{table}[h]
  \caption{Comparison of the thermal activation energy E$_A$ and the characteristic temperature T$_C$ of the various structures.}
  \begin{tabular}[htbp]{@{}llllll@{}}
    \hline
     Structure & (a)  & (b) & (c) & (d) & (e)  \\
    \hline
    E$_A$ (meV) & 150  & 205 & 220 & 251 & 272 \\
    T$_C$ (K)   & 150  & 140 & 150 & 160 & 192 \\
    \hline
  \end{tabular}
  \label{comp}
\end{table}
Within the error the thermal activation energies can be increased from E$_A$= 150~meV for the reference structure to about 270~meV for the cladding structure with tunnel barriers. This increase can be attributed to the influence of the additional band offset of the barrier configurations. Furthermore, the charactersitic temperatures significantly increase from 150~K up to 192~K. The reason for the low T$_C$-value of structure (b) is currently now not known.

\subsection{Micro-PL and autocorrelation measurements}
\begin{figure}[h]
\includegraphics[width=01\linewidth]{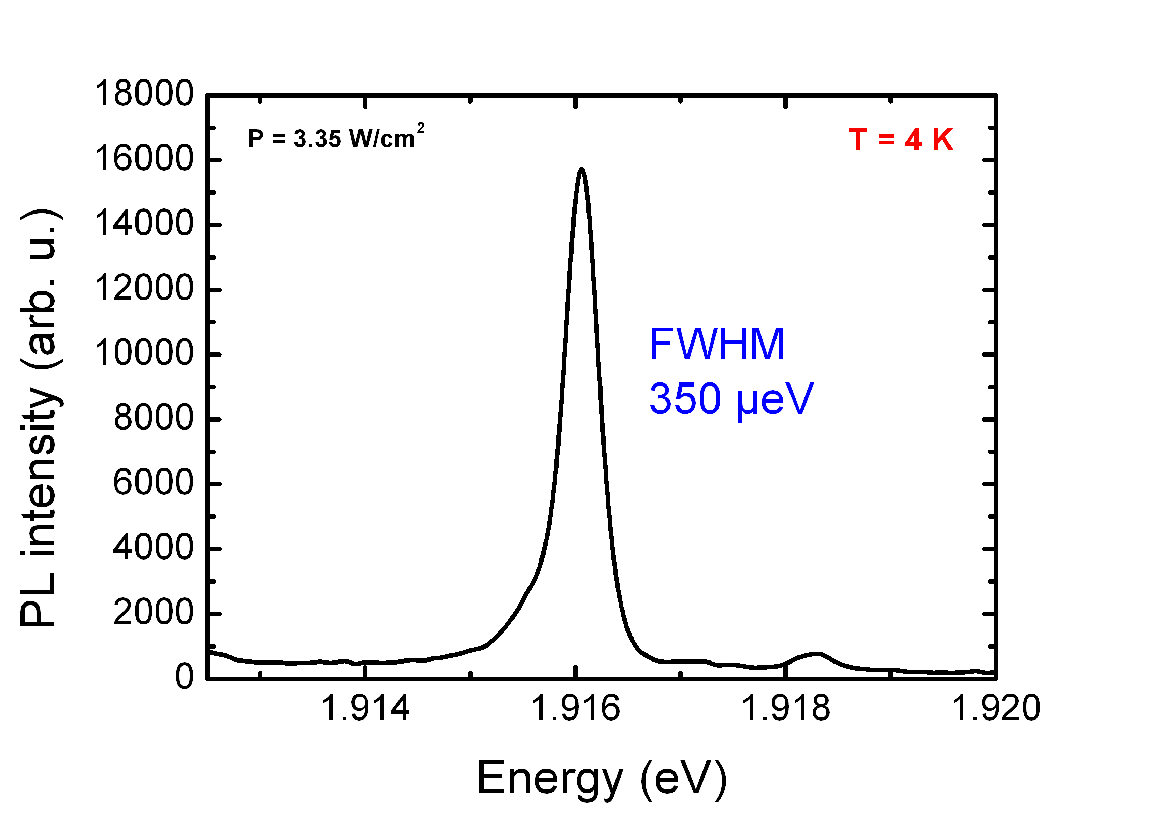}
\caption{$\mu$-PL measurement of a single InP-QD in an Al$_{0.20}$GaInP-barrier. For single-QD spectroscopy the sample was structured using a nanosphere lithography process.} \label{M4090_Single}
\end{figure}

In addition to the ensemble measurements, the zero-di\-men\-sio\-nal\-ity of the dots was confirmed by performing $\mu$-PL and second-order autocorrelation measurements on the samples. For single QD spectroscopy a 50 nm Cr layer was deposited and structured using a nanosphere lithography process (NSL) to produce holes with diameters of around 500~nm \cite{hakanson2008}. Fig. \ref{M4090_Single} displays the $\mu$-PL spectra of the emission of a single QD with the reference Al$_{0.20}$GaInP-barrier, emitting at 1.916~eV at 4~K. The relatively large linewidth of 350~$\mu$eV at 4~K is due to local carrier density fluctuations, also known as spectral diffusion, in vicinity of the QD \cite{wang04}.

The zero-dimensional nature of the emission source can be observed in the second-order autocorrelation measurement g$^2$($\tau$) of the QD emission line at 4~K.

\begin{figure}[h]
\includegraphics[width=1\linewidth]{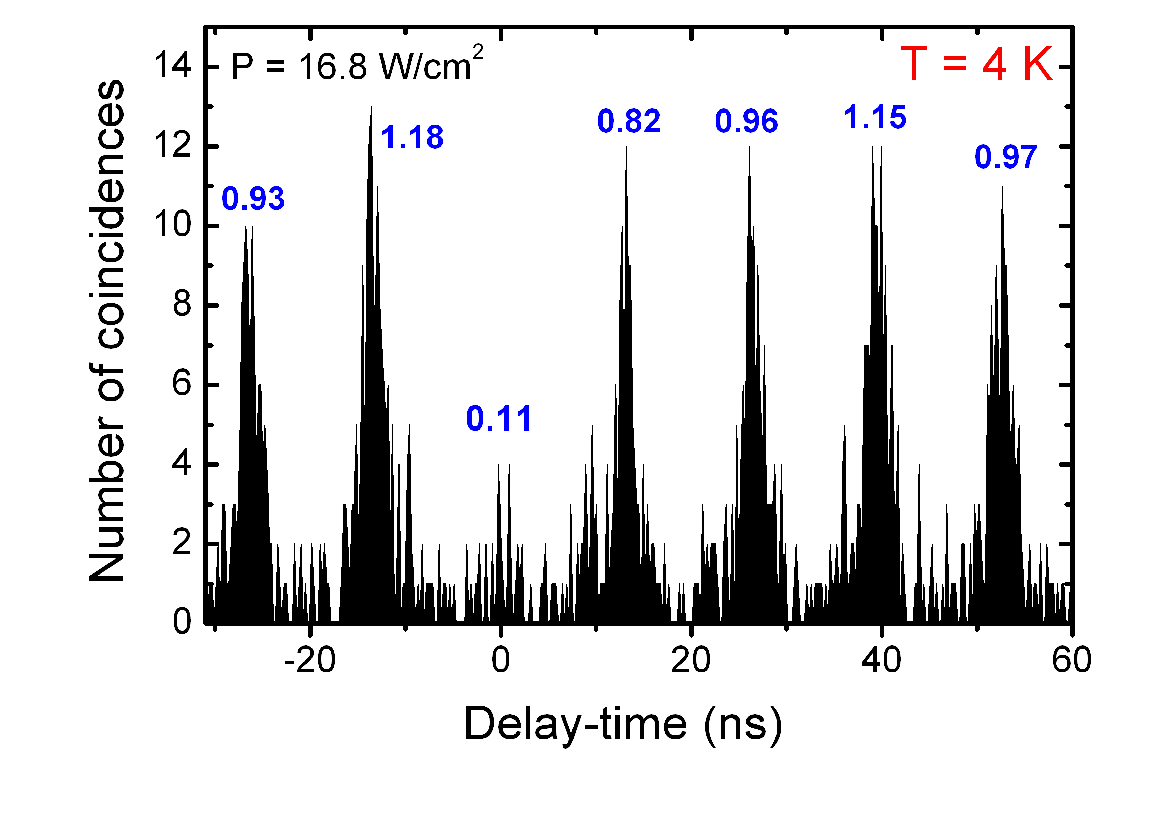}
\caption{Autocorrelation measurement of the single-QD shown in Fig. \ref{M4090_Single} which displays a clear single-photon emission behavior at 4~K.} \label{M4090_auto_4K}
\end{figure}

The suppression of coincidence events at $\tau$=0 is equivalent to 11$\%$ of the calculated Poisson-normalized level (Fig. \ref{M4090_auto_4K}). This observation in turn indicates that the luminescence originates from a single InP-QD and that our structures are therefore capable of providing triggered single-photon emission.

\section{Conclusion}
To summarize, in this work we have shown that the use of different barrier configurations does not affect the growth and therefore the optical characteristics of the investigated QDs. The temperature stability and the relative PL intensity can be strongly increased, using very simple barrier configurations (tunnel barriers) or more complex structures (combination of tunnel barriers with abrupt claddings). With the intensity from the latter increased by a factor of up to 20 at 240~K with respect to structures without barrier claddings.
In addition, the estimated activation energy can be increased by increasing the total band offset of the cladding.
Barrier claddings are therefore highly recommended for applications at elevated temperatures, such as p-i-n-LED and laser structures. In addition, $\mu$-PL and autocorrelation experiments revealed the zero-dimensional nature of the QDs by clearly showing single-photon emission (antibunching) characteristics.

\begin{acknowledgement}
Financial support by SFB TRR 21 and FOR 730 is gratefully acknowledged. We also would like to thank E. Kohler for technical assistence with the MOVPE system. 
\end{acknowledgement}

%
%

\end{document}